\documentclass[fleqn]{ILASSEurope_2025_Paper_Format}
\usepackage[version=4]{mhchem}
\usepackage{graphicx}
\usepackage{subcaption,ulem} 
\usepackage{booktabs}

\begin{document}
\title{Reactive Phase Change of a Single Droplet in Turbulent Flow with Reactive Vapor--Liquid Equilibrium Using a Two-Scalar Diffusive-Interface Method}

\doi{}
\vspace{1mm}
\author{Sajad Jafari$^{*\text{1}}$, Silvia Lasala$^{\text{2}}$, Ran Yao$^{\text{1}}$, Christophe Duwig$^{\text{1}}$}
\vskip 5mm
\address{$^{\text{1}}$ Department of Chemical Engineering, Royal Institute of Technology (KTH), Sweden\\
$^{\text{2}}$ Université de Lorraine, CNRS, LRGP, F-54000 Nancy, France\\
\vskip 2pt
{\normalsize{*Corresponding author:
\href{mailto:sajadja@kth.se}{\textstyleInternetlink{sajadja@kth.se}}}}}
\vspace{5mm}
\section*{Abstract}
Reactive phase change fluids have been shown to enhance thermodynamic performance by integrating chemical reactions within thermodynamic cycles. This study investigates the turbulent evaporation of a single droplet using Direct Numerical Simulation (DNS) to model the behavior of reactive phase change fluids. Applying Reactive vapor-liquid equilibrium (RVLE) thermodynamics, the study simultaneously captures the dynamics of reactive phase transitions, mass transfer, and heat transfer during the evaporation of dinitrogen tetroxide (N$_2$O$_4$). 
A second-order phase field framework is used to derive two scalar transport equations, accurately describing species behavior at equilibrium. The equilibrium is maintained by balancing concentration exchange rates across both the liquid-to-vapor and vapor-to-liquid phases. Results indicate that chemical reactivity significantly influences phase transition dynamics, altering the phase dome structure by modifying critical points and the thermodynamic saturation lines. 
Endothermic and exothermic behaviors are observed at the interface during the fast reversible evaporation process of 
N$_2$O$_4$(v,l) $\rightleftharpoons$ 2NO$_2$(v,l). 
Turbulence simulations further show enhanced mixing efficiency and improved gas-liquid separation, driven by chaotic flow patterns. This study elucidates the role of reactive phase change fluids in contrast to conventional non-reactive phase flows, emphasizing the influence of chemical reactions on phase transition dynamics and overall system performance.

\vspace{5mm}
\textbf{Keywords:} DNS, Reactive Vapor-Liquid Equilibrium, Reactive Evaporation, Diffuse Interface, Two scalar  
\section{Introduction}
Optimizing power generation efficiency is essential for achieving sustainable energy solutions, as current heat engines waste approximately 75\% of global primary energy, much of which is in the form of low-grade heat (\( T < 100^\circ \mathrm{C} \)) \cite{IRENA2019}. To mitigate this, there is increasing interest in utilizing reactive fluids, which have the potential to symbiotically convert thermal and chemical energy, thereby enhancing efficiency \cite{ran2025,lasala2021, lasala2022}. In this context, our recent computational fluid dynamics (CFD) studies have demonstrated that the application of reactive fluids in single-gaseous-phase systems can significantly enhance heat transfer coefficients—by up to 600\%\cite{ran2025, Zhang2024}. This improvement is attributed to the combined effects of forward and backward reactions, which intensify local temperature gradients, leading to better heat exchange. Furthermore, our thermodynamic investigations on two-phase reactive systems \cite{lasala2024} provide additional evidence supporting this enhancement within the framework of reactive vapor-liquid equilibrium. At the forefront of this exploration is the concept of RVLE, which elucidates the intricate interplay between chemical reactions and phase transitions \cite{barbosa1987, lasala2024}. The presence of rapid and reversible chemical reactions further complicates this landscape, particularly when coupled with the diverse physics occurring across multiple temporal and spatial scales \cite{lasala2024}. Therefore, robust modeling approaches are crucial for capturing these intricate processes in a detailed, thermodynamically, and mathematically consistent manner.

\begin{figure}[t!]
    \vspace{4pt}
    \centering
    \includegraphics[scale=.6]{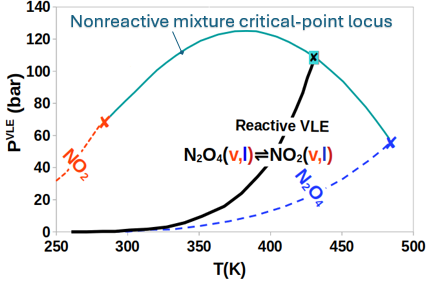}
\caption{P-T diagram for the $\text{N}_2\text{O}_4$-$\text{NO}_2$ system (Peng-Robinson EoS): phase diagrams of pure NO$_2$ and N$_2$O$_4$ (dashed) and the reactive system (black), along with the critical locus of the non-reactive mixture (cyan)~\cite{lasala2024}.}
    \label{fig:PT diagram}
\end{figure}

From a thermodynamic perspective, integrating chemical reactions into vapor-liquid equilibrium alters phase behavior significantly \cite{barbosa1987, lasala2024}. A key example is the dinitrogen tetroxide (N$_2$O$_4$) system, where rapid and reversible dissociation into nitrogen dioxide (NO$_2$) ($\mathrm{N}_2\mathrm{O}_4\mathrm \leftrightarrow \mathrm{2NO}_2$) modifies the phase equilibrium, as shown in Fig.~\ref{fig:PT diagram} \cite{lasala2024}.

From a mathematical perspective as well, capturing interfacial behavior in two-phase flows, where distinct fluid phases coexist, remains a significant challenge, particularly at micro- to nanoscale regimes. Modeling the two-phase flows requires precisely capturing the interface dynamics between phases, which is typically achieved using either sharp or diffuse interface methods (DIM)~\cite{Mirjalili2019, sal2024, Scapin2022}. For the present study, the recently developed DIM approach is adopted, which effectively captures the complex dynamics of interfacial flow, enabling a more comprehensive analysis of the reactions occurring at the phase interface \cite{Mirjalili2022}. Among the various DIM models, those that assume equilibrium in momentum, temperature, and pressure across the phases—while treating the interface as a kinetically controlled nonequilibrium process—offer an effective balance between computational efficiency and model complexity~\cite{yip, Jafari2022}. These models enhance computational efficiency by employing averaged transport equations, where the conservation of momentum and enthalpy is described through single-scalar formulations, while species diffusion is governed by two scalar equations within each phase. This strikes an effective balance between accuracy and computational feasibility, making it well-suited for analyzing interfacial reactive mass transfer in multiphase systems.

\textbf{To outline briefly}, this study presents a newly developed modeling tool to explore the complex dynamics of turbulent reactive phase change through reactive VLE thermodynamics, with a focus on a single N$_2$O$_4$ droplet surrounded by a turbulent NO$_2$-N$_2$O$_4$ gas phase. Direct numerical simulation is employed to solve the two-scalar transport equations, which describe the species behavior at the reactive liquid-vapor interface while incorporating equilibrium concentration exchange rates during phase transitions. These transport equations are coupled with single-scalar equations governing momentum and enthalpy conservation, and the system is closed by a second-order phase-field equation that governs the dynamics of the phase boundary and interface evolution. To account for non-ideal VLE effects, the model introduces a locally varying apparent Henry's constant, which depends on thermodynamic variables, ensuring thermodynamic consistency in phase interactions. This approach is based on our recently validated methodology, enabling the assessment of reactive VLE \cite{lasala2024}. The formulations capture interfacial reactive heat and mass transfer within the diffuse interface framework, including Stefan flow effects, providing insight into the coupling between turbulence, phase change, and reactive transport in multiphase systems.

 \section{Mathematical Model}\label{MathematicalModel}

The following assumptions have been considered.
\begin{itemize}
    \item The gas phase is assumed to behave as an ideal gas, which is valid under the ambient conditions of low temperature and pressure.

    \item Constant thermal and transport properties are assumed for individual components in the gas and liquid phases, due to the limited variation in pressure.
    
    \item The properties of the liquid, gas, and two-phase mixtures are computed using a linear mixing rule, which incorporates the composition and phase fractions of the individual components.

    \item At the droplet interface, thermo-chemical equilibrium is assumed, and the calculation of the equilibrium concentration for both gas and liquid is performed using a locally variable Henry's constant.
    
  \item All reactions are modeled exclusively in the gas phase.      
  
  \item Gravity effects are ignored.

\end{itemize}

\subsection{Governing equations}
\subsubsection{Phase Field Equation}
A phase-field model introduced by~\cite{Mirjalili2022} is used to define the spatial distribution of the liquid, gas, and diffused interface regions. The phase field parameter $\phi$ indicates the phases, where $\phi = 1$ for the liquid, $\phi = 0$ for the gas, and intermediate values ($0 < \phi < 1$) represent the interface region.

\begin{equation}
\label{eq:model_equations}
    \frac{\partial \phi}{\partial t} + \nabla \cdot (\vec{u} \phi) + \frac{\dot{m_{\Gamma}}}{\rho_l} = \nabla \cdot \left\{\gamma \left[\epsilon \nabla \phi - \phi (1 - \phi) \frac{\nabla \phi}{\left|\nabla \phi\right|} \right]\right\}
\end{equation}

In this context, \(\epsilon\) serves as the parameter controlling the thickness of the diffusive interface and is assumed to be approximately close to the mesh size, here \(\epsilon = \Delta x\). The parameter \(\gamma\) in the model is determined based on the maximum absolute velocity within the domain, as proposed by~\cite{chiu2011}. Additionally, $u_\Gamma$ represents the non-divergence-free interface velocity, which is used in the advection term, as detailed later.

\subsubsection{Two-Scalar Transport for Concentration}
Given the focus on individual species behavior, including phase transitions and chemical reactions, we employ two-scalar transport equations in molar concentration form for both the liquid and gas phases~\cite{Mirjalili2022,sal2024}, as given in Eq.s~\ref{eq:two_scalar1}-\ref{eq:two_scalar2}.

\begin{align}
\label{eq:two_scalar1}
\frac{\partial c_{1,i}}{\partial t} + \nabla \cdot (\vec{u} c_{1,i}) &= \nabla \cdot \left[ D_{1,i} \left( \nabla c_{1,i} - \left(\frac{1-\phi}{\epsilon}\right) \vec{n} c_{1,i}  \right) \right]- S_{12,i}  \\
\label{eq:two_scalar2}
\frac{\partial c_{2,i}}{\partial t} + \nabla \cdot (\vec{u} c_{2,i}) &= \nabla \cdot \left[ D_{2,i} \left( \nabla c_{2,i} + \frac{\phi}{\epsilon} \vec{n} c_{2,i} \right) \right] + S_{12,i} + R_{2,i}
\end{align}

Where indices 1 and 2 correspond to the liquid and gas phases, respectively, \( i \) represents the species index within each phase, and \( \vec{n} = \frac{\nabla \phi}{|\nabla \phi|} \) is the normal vector. The concentration \( c \) is expressed in \(\text{mol/m}^3\), and \( D \) denotes the diffusion coefficient (\(\text{m}^2/\text{s}\)). The term \( S_{12} \) represents the interfacial mass transfer, while \( R_{2} \) denotes the reaction source term in the gas phase.

\paragraph{Mass transfer mechanism}  
\mbox{}  
\label{sec:Mass transfer Scheme}

The interfacial mass transfer term, \( S_{12,i} \), in Eq.s~\ref{eq:two_scalar1}-\ref{eq:two_scalar2} represents the net flux of species \( i \) from phase 1 (liquid) to phase 2 (gas). At equilibrium, the rates of concentration exchange for both the liquid-to-gas and gas-to-liquid processes equalize, resulting in no net transfer between phases, such that \( S_{12,i} + S_{21,i} = 0 \) for each species \( i \). This condition reflects mass conservation and ensures constant concentrations of species within the system over time. The interfacial transfer term for species \( i \) is given by Eq.~\ref{eq:SJ12}.
\begin{equation}
S_{12,i} = -A D_{12,i} \left[ E_{12,i} \, c_{2,i} \, \phi - c_{1,i} \, (1 - \phi) \right] + D_{12,i} \nabla \phi \cdot \nabla \left( c_{1,i} + E_{12,i} \, c_{2,i} \right)
\label{eq:SJ12}
\end{equation}
Where \( S_{12,i} \) for each species is governed by the concentrations (\( c_{1,i} \), \( c_{2,i} \)) in both phases, the phase field parameter \( \phi \), the equilibrium concentration ratio \( E_{12,i} \), and the interfacial diffusion coefficient \( D_{12,i} \), as expressed in Eq.~\ref{eq:D12}. The parameter \( A \) represents the inverse time scale for equilibrium and is typically chosen as a large constant (e.g., \( A = 1000 \)) without imposing additional time-step restrictions \cite{Mirjalili2022}.
\begin{equation}
D_{12,i} = \frac{D_{1,i} \, D_{2,i}}{E_{12,i} \, D_{1,i} \, (1 - \phi) + D_{2,i} \, \phi}
\label{eq:D12}
\end{equation}
The equilibrium concentration ratio, \( E_{12,i} \), defines the phase equilibrium condition for a species at the interface, as given in Eq.~\ref{eq:Keq}. It is modeled using the RVLE thermodynamic solver~\cite{lasala2024}, as shown in Fig.~\ref{fig:PT diagram}.
\begin{equation}
E_{12,i} = H_{i}(T,P) = \frac{X_{l,i}}{Y_{g,i}} \quad \text{for species i}
\label{eq:Keq}
\end{equation}
\paragraph{Chemical Reaction Mechanism and Modeling}  
\mbox{}  
\label{sec:Reaction Scheme}

The reversible reaction \( R_{2,i}=\nu_{n_{(p,r)}} \dot{r}_n \) follows the mass action law, where the reaction rate is given by \( \dot{r} = k^f \prod_{nr \in \text{react}} [c_{nr}]^{-\nu_{nr}} - k^r \prod_{np \in \text{prod}} [c_{np}]^{\nu_{np}} \). Here, \( k^f \) and \( k^r = \frac{k_f}{K_{\text{eq}}} \) represent the forward and backward rate constants, respectively, while \( \nu_{np} \) and \( \nu_{nr} \) are the stoichiometric coefficients for the product and reactant. \( K_{\text{eq}} \) is the reaction equilibrium constant, calculated using the Ulich approximation as \( \ln K_{\text{eq}}(T) = -\frac{\Delta_R G^\circ(T)}{RT} = -\frac{\Delta_R H^\circ(T_0)}{RT} + \frac{\Delta_R S^\circ(T_0)}{R} \) \cite{lasala2024}. $T_0$ is an arbitrary temperature, where it is set here to $T_0 = 298$~K. The standard molar enthalpies \( \Delta_R H^\circ(T_0) \) and standard molar entropies \( \Delta_R S^\circ(T_0) \) for various species are summarized in Table~\ref{tab:Heat_of_Formation} \cite{nist_therm}. The forward rate $k_f$ is taken from literature~\cite{Borrell1988, nist2}.
\begin{table}[ht]
    \centering
\caption{Table~\ref{tab:Heat_of_Formation}: Standard molar enthalpy of formation and standard molar entropy at 298.15 K, \cite{nist_therm}.}
    \label{tab:Heat_of_Formation}
    \begin{tabular}{cccc}
        \hline 
        Phase & Species & $ H_{f,298}^{\circ}$ (J/mol)& $ S_{f,298}^{\circ} (J/mol.k)$ \\
        \hline
        gas & $N_2O_4$ & 9079 & 304.3 \\
         \hdashline
         gas & $NO_2$ & 33180 & 239.9
    \end{tabular}
\end{table}
\vspace{-5mm}
\subsubsection{{One-Scalar transport for Enthalpy}}\
Predicting temperature distributions is essential to ensure heat conservation during interfacial reactive mass transfer. The enthalpy equation is expressed as:
\begin{subequations}
\begin{align}
&\frac{\partial \rho h}{\partial t} + \nabla \cdot \left( \vec{u} (\rho h) \right) = \nabla \cdot (k  \nabla T) + h_{rxn} - \dot{m}_{\Gamma} h_{lv} \delta_{\Gamma} \label{eq:enthalpy} 
\end{align}
\end{subequations}
In this context, \( \rho \) denotes mixture density, \( k \) thermal conductivity, \( h = c_p T \) is specific enthalpy (with temperature \( T \) and specific heat capacity \( c_p \)), and \( h_{lv} \) represents latent heat for phase transitions. Mixture properties are computed using a linear mixing rule considering component compositions and phase fractions. The thermodynamic properties of individual species, as listed in Table~\ref{tab:Thermophysical}, are computed using our thermodynamic solver with the Peng-Robinson (PR) EoS \cite{lasala2024}. The heat reaction rate \( h_{rxn} \) is given by \(h_{rxn} = -\dot{r} \left( \sum \nu_{np} h_{f^{\circ},np} - \sum \nu_{nr} h_{f^{\circ},nr} \right) \). Here,  viscous dissipation is neglected in the enthalpy equation due to its negligible contribution under the low Mach and moderate Reynolds number conditions, which is standard in DNS of interfacial heat transfer.

\begin{table}[h!]
    \centering
    \caption{Thermophysical properties of the two-phase system calculated using the PR EoS as an extension to mixtures, based on the PhD thesis of Dehlouz A.~\cite{Dehlouz2022}.}
    \label{tab:Thermophysical} 
     \small 
    \begin{tabular}{cccccccc}      
        \hline
        Phase & T(K) & Species & $\mu$ [kg/(m$\cdot$s)] & $c_p$ [J/(kg$\cdot$K)] & $\kappa$ [W/(m$\cdot$K)] & $D$ [m$^2$/s] & c [mol/m$^3$] \\
        \hline      
        Liquid & 270 & N$_2$O$_4$ & $6.02 \times 10^{-4}$ & 1502 & 0.1588 & $2.02 \times 10^{-9}$ & 15,000 \\
        \hdashline  
        Gas & 352 & N$_2$O$_4$ & $1.41 \times 10^{-5}$ & 910.68 & 0.0149 & $1.05 \times 10^{-6}$ & 374 \\
        Gas & 352 & NO$_2$ & $1.63 \times 10^{-5}$ & 838.73 & 0.0179 & $3.01 \times 10^{-6}$ & 617 \\
    \end{tabular}
\end{table}
\vspace{-10pt}
\subsubsection{{One-Scalar transport for Momentum}}
The momentum equation is solved using a conventional pressure correction method, as shown in Eq.s \ref{eq:momentum}-\ref{eq:final_velocity_update},  with detailed information provided in~\cite{Scapin2022}.
\begin{align}
&\frac{\partial (\rho \vec{u})}{\partial t} + \nabla \cdot (\vec{u} (\rho \vec{u} )) = -\nabla \mathbf{P} + \nabla \cdot \left[ \mu \left( \nabla \vec{u} + \nabla \vec{u}^\top \right) \right]+f_{x_{ijk}}+f_{\sigma} \label{eq:momentum} \\
&\frac{\vec{u}^{**} - \vec{u}^{n}}{\Delta t^{n+1}} = f_{t,1}\textbf{R}^n + f_{t,2}\textbf{R}^{n-1} \label{eq:momentum_update} \\
&\vec{u}^{*} = \vec{u}^{**} - \frac{\Delta t^{n+1}}{\rho_0^{n+1}} \left[\left(1-\frac{\rho_0}{\rho^{n+1}}\right) \nabla \hat{p} + \nabla p^n\right] \label{eq:velocity_update} \\
&\nabla^2 \delta p^{n+1} = \frac{\rho_0^{n+1}}{\Delta t} (\nabla \cdot \vec{u}^{*} - \nabla \cdot \vec{u}_\Gamma) \label{eq:pressure_poisson} \\
&p^{n+1}=p^{n}+\delta p^{n+1}\label{eq:final_pressure_update}\\ 
&\vec{u}^{n+1} = \vec{u}^{*} - \frac{\Delta t^{n+1}}{\rho_0} \nabla \delta p^{n+1} \label{eq:final_velocity_update} 
\end{align}
\( f_{\sigma} \) represents the surface tension force induced by the Marangoni effect\cite{Scapin2020}. Interfacial reactive mass transfer induces a velocity discontinuity, resulting in a nonzero velocity divergence at the interface. To account for this effect, the velocity field is formulated by incorporating the influence of reactive mass flux, as described in Eq.s \ref{eq:mass flux} and \ref{eq:mass flux2}.
\begin{flalign}
\nabla \cdot \vec{u}_{\Gamma} &=  \dot{m}_{\Gamma} \left(\frac{1}{\rho_{g,\Gamma}} - \frac{1}{\rho_{l,\Gamma}}\right)\delta_{\Gamma} \label{eq:mass flux} \\
\dot{m_{\Gamma}} &= \sum_{i=1}^{n} \dot{m}^i_{{l,g}} = -\sum_{i=1}^{n} D^i_{12} \nabla c^i_{12} M_{w,i} \vec{n_\Gamma}  \label{eq:mass flux2}
\end{flalign}
\( \nabla \cdot \vec{u}_{\Gamma} \) and \( \dot{m}_{\Gamma} \) represent the interfacial velocity divergence and interfacial mass flux. \( \delta_{\Gamma} \) is the Dirac delta function, localizing the mass flux at the interface, and \( \vec{n}_\Gamma \) is the normal vector at the interface. \( D_{12,i} \) is the mixture diffusion coefficient, \(M_{w,i}\) is the molecular weight, and \(\nabla c^i_{12}\) denotes the concentration gradient at the interface, calculated through the scalar equations, for species \(i\).
\subsection{Direct Numerical Simulation (DNS)}
We employ the Arnold-Beltrami-Childress (ABC) method to simulate turbulence within the framework of DNS, incorporating it as a source term in the momentum equation (Eq.~\ref{eq:momentum}). This method governs the evolution of sinusoidal wave modes, as described in Eq.~\ref{eq:ABC} \cite{arnold2014}.
\begin{equation}
    \begin{aligned}
        \label{eq:ABC}
        f_x &=\rho u_0  \left( A \sin(k_0 z) + C \cos(k_0 y) \right)(1-\phi) \\
        f_y &= \rho u_0  \left( B \sin(k_0 x) + A \cos(k_0 z) \right) (1-\phi)\\
        f_z &= \rho u_0 \left( C \sin(k_0 y) + B \cos(k_0 x) \right) (1-\phi)
    \end{aligned}
\end{equation}

Where \( u_0 \) and \( k_0 \) represent the turbulence flow amplitude and the wave number, respectively, and \( A \), \( B \), and \( C \) control the amplitudes of the sinusoidal components in various directions. Setting \(A = B = C = 1\) yields uniform stirring and symmetric flow. 
\section{Results}
\label{sec:results}
Simulations are initialized with an N\(_2\)O\(_4\) droplet at \(T_L = 270 \, \text{K}\) and an equilibrium gas mixture at \(T = 352 \, \text{K}\), with initial conditions as summarized in Table \ref{tab:Thermophysical}. The geometry configuration used in the simulations is detailed in Table~\ref{tab:Geometry case setup}.
\vspace{-10pt}
\begin{table}[htbp]
    \centering
\caption{Geometry configuration for the droplet and the cubic box, with all dimensions presented in mm. The variables $r_{c}$, $L_{b}$, $n_{L_i}$, and $\Delta x_{i}$, $n_{Di}$ denote the droplet radius, box length, number of grids, mesh size, and number of grids per droplet along each axis, respectively.}
    \label{tab:Geometry case setup}    
    \begin{tabular}{ccccc} 
        \hline       
     $r_{c}$ & $L_{b}$ & $n_{L_i}$ & $\Delta x_{i}$ &$n_{Di}$\\      \hline     
       0.2 &1.5&256&0.006 &68
    \end{tabular}
\end{table}
\vspace{-10pt}

\begin{figure}[b!]
    \centering
    \begin{subfigure}[b]{0.45\textwidth}
        \centering
        \includegraphics[scale=0.325]{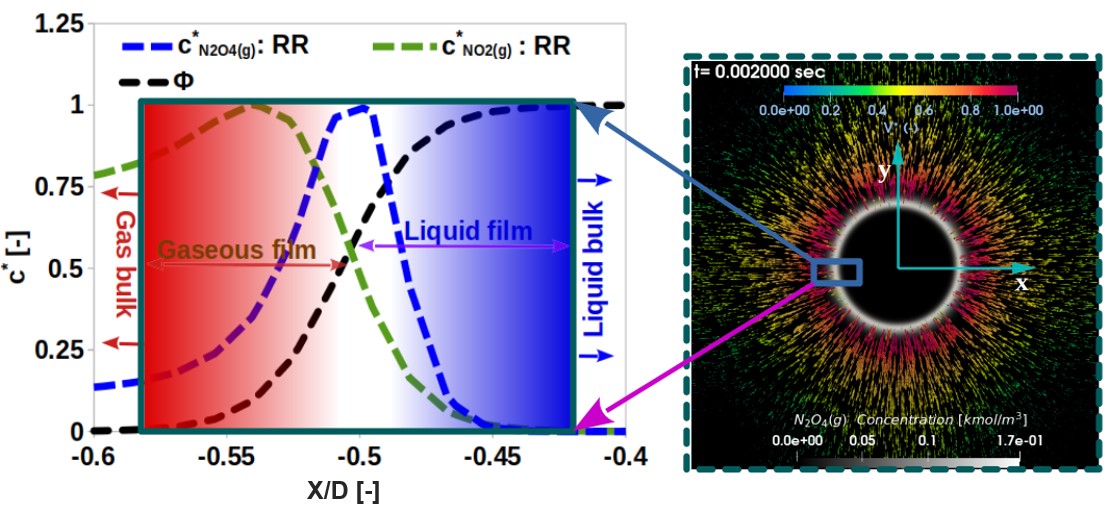}
        \caption{RVLE interface} 
        \label{FIG:Stefan}
    \end{subfigure}
    \hfill
    \begin{subfigure}[b]{0.44\textwidth}
        \centering
        \includegraphics[scale=0.305]{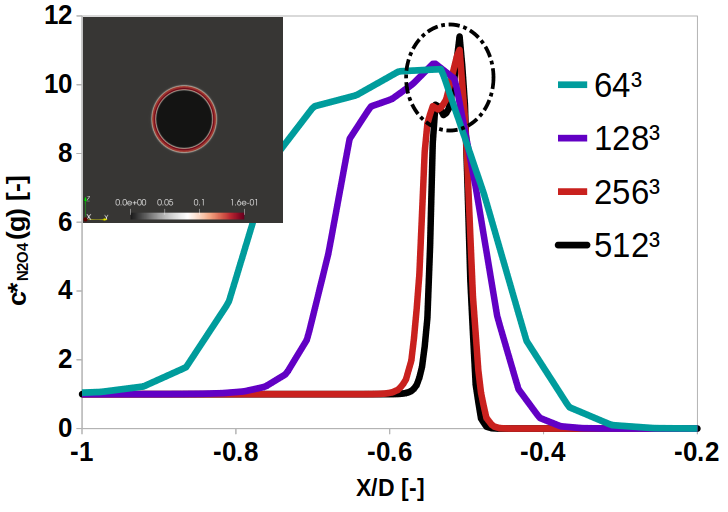}
        \caption{Grid independency}
        \label{FIG:Grid}
    \end{subfigure}
\caption{(a) RVLE within a two-scalar diffuse interface under the influence of a nonzero interfacial gradient velocity, and (b) grid independence analysis of the normalized concentration of reactive evaporated \( \text{N}_2\text{O}_4(g) \), defined as \( C^*_{\text{N}_2\text{O}_4(g)} = \frac{C_{\text{N}_2\text{O}_4(g)}}{C^{\text{init}}_{\text{N}_2\text{O}_4(g)}} \), across varying grid resolutions.}
\label{fig:combined}
\end{figure}

To elucidate the interfacial dynamics within the two-scalar framework, Figure~\ref{fig:combined}(a) presents the spatial distribution of the reactive species and velocity field at the laminar interface. The phase field parameter \(\phi\) transitions smoothly from 0 (gas phase) to 1 (liquid bulk), defining the diffuse interfacial region where both phases coexist. This transition governs the interfacial mass transfer and reaction dynamics.  

Figure~\ref{fig:combined}(a) further illustrates the dimensionless concentration profiles of evaporated \(\text{N}_2\text{O}_4\) and its reactive product \(\text{NO}_2\)(g), resolved along the interface. Reactive mass transfer is governed by both gas and liquid phase transport, with bidirectional species interactions, resulting in an asymmetric distribution of species concentrations at the interface. Overlaid vector fields, colored by velocity magnitude, highlight the presence of an interfacial reactive gradient velocity, reinforcing the coupled nature of reaction and mass transfer. Additionally, to further investigate the impact of the diffuse interface at the phase boundary, a grid independence study was performed, as shown in Figure~\ref{FIG:Grid}. The reactive evaporated \(\text{N}_2\text{O}_4\) concentration was evaluated at different grid resolutions ranging from \(64^3\) to \(512^3\). While the overall concentration magnitude remained consistent across grid resolutions, diffusion behavior varied significantly. Coarser grids exhibited excessive numerical diffusion, leading to broader concentration profiles, whereas finer grids preserved sharper gradients at the interface. These studies confirm the necessity of sufficient grid resolution to capture interfacial dynamics accurately without introducing artificial diffusion effects.
\begin{figure*}[t!]
  \centering
  \begin{subfigure}{0.37\linewidth}  
    \includegraphics[width=\linewidth]{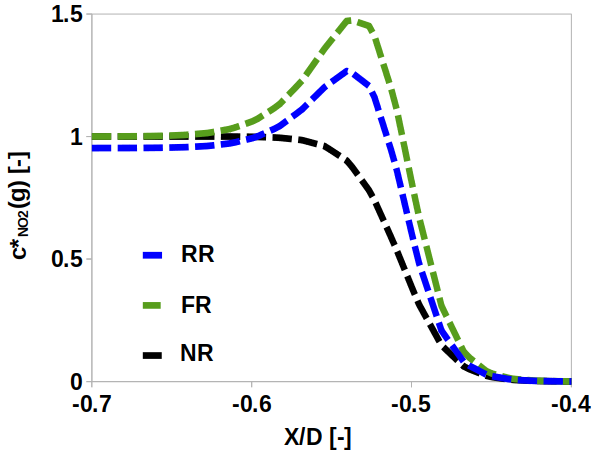}
\caption{\(C^*_{\text{NO}_2(g)} = \frac{C_{\text{NO}_2(g)}}{C^{\text{NR, init}}_{\text{NO}_2(g)}}\)}
    \label{fig:no2_conc_var_no2FRN}
  \end{subfigure}
  \hspace{0.04\linewidth}  
  \begin{subfigure}{0.37\linewidth}  
    \includegraphics[width=\linewidth]{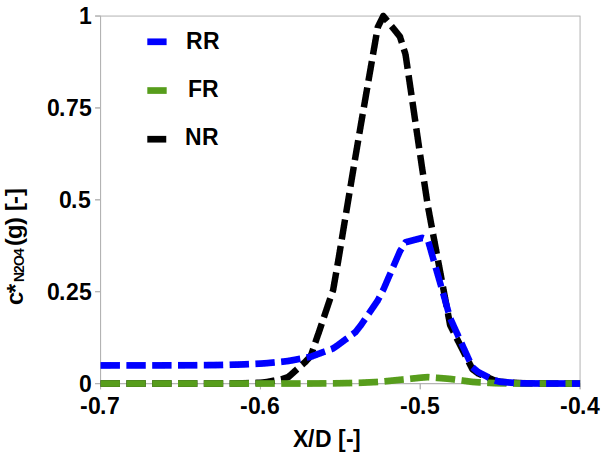}
\caption{\(C^*_{\text{N}_2\text{O}_4(g)} = \frac{C_{\text{N}_2\text{O}_4(g)}}{C^{\text{NR, max}}_{\text{N}_2\text{O}_4(g)}}\)}   
    \label{fig:n2o4_conc_var_no2}
  \end{subfigure}

\caption{Spatial distribution of normalized (a) \(\text{NO}_2(\text{g})\) and (b) evaporated \(\text{N}_2\text{O}_4\) concentrations under different reaction mechanisms (NR: Non-Reactive, FR: Forward Reaction, RR: Reversible Reaction).}
  \label{fig:liq_conc}
\end{figure*}

The different reaction mechanisms and their associated reactivity also influence the topology of the interface, with each mechanism impacting the system in distinct ways. Figure \ref{fig:liq_conc} shows the normalized concentration distributions for the different mechanisms (NR: Non-Reactive, FR: Forward Reaction, RR: Reversible Reaction). The Non-Reactive flow (black dashed line) leads to the highest concentration of \(\text{N}_2\text{O}_4\) at the interface, with no \(\text{NO}_2\) formation. In the Forward Reaction (red dashed line), \(\text{N}_2\text{O}_4\) is consumed, and \(\text{NO}_2\) is produced, which reduces the interfacial \(\text{N}_2\text{O}_4\) concentration. The Reversible Reaction (blue dashed line) shows a balanced concentration distribution of both \(\text{N}_2\text{O}_4\) and \(\text{NO}_2\), driven by both the forward and backward reactions, which affect the concentration gradients at the interface.
This analysis is crucial for understanding the dynamics of interfacial reactive heat transfer, governed by the heat reaction rate \( h_{\text{rxn}} \) within the reversible reaction mechanism. The phase change, influenced directly by the liquid concentration of \( \text{N}_2\text{O}_4 \), impacts the production of gaseous \( \text{N}_2\text{O}_4 \) and, in turn, affects the reaction rate, governed by \( \dot{r} = k_f(T) \left( c_{\text{N}_2\text{O}_4\text{ (g)}} - \frac{1}{K_{\text{eq}}(T)} c_{\text{NO}_2\text{ (g)}}^2 \right) \).

\begin{figure*}[b!]
  \centering
  \begin{subfigure}{0.46\linewidth}
    \includegraphics[width=\linewidth]{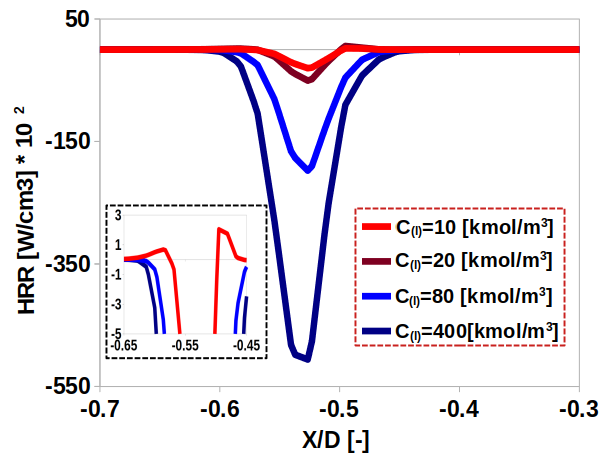}
  \end{subfigure}
\caption{Spatial distribution of the Heat Reaction Rate (\(HRR\)) for different initial concentrations of liquid \(\text{N}_2\text{O}_4\) ranging from 10 to 400 kmol/m\(^3\), at a fixed initial gaseous \(\text{NO}_2\) concentration.}  
\label{fig:TEMP_ENTHALPI}
\end{figure*}
As shown in Fig. \ref{fig:TEMP_ENTHALPI}, the distribution of the Heat Reaction Rate (HRR) illustrates how variations in liquid \(\text{N}_2\text{O}_4\) concentration affect the enthalpy balance and modify the reactive system's thermal behavior. The results reveal a local transition from exothermic to endothermic behavior as the coupling between phase change and the forward reaction intensifies. The phase change of liquid \( \text{N}_2\text{O}_4 \) alters the gas-phase concentration of \( \text{N}_2\text{O}_4 \), which affects the reaction rate \( \dot{r} \), governed by the concentrations of \( \text{N}_2\text{O}_4(g) \) (forward reaction) and \( \text{NO}_2(g) \) (reverse reaction). As the liquid \( \text{N}_2\text{O}_4 \) concentration increases, the forward reaction suppresses the backward reaction, reducing exothermic heat release and possibly shifting the system towards a more endothermic state. Hence, the thermal dynamics are governed by the interaction of phase change, reaction kinetics, and heat flux, as depicted in Fig. \ref{fig:TEMP_ENTHALPI}.
\begin{figure}[t!]
    \centering
    \begin{subfigure}{0.55\textwidth}
        \centering
        \includegraphics[width=\textwidth]{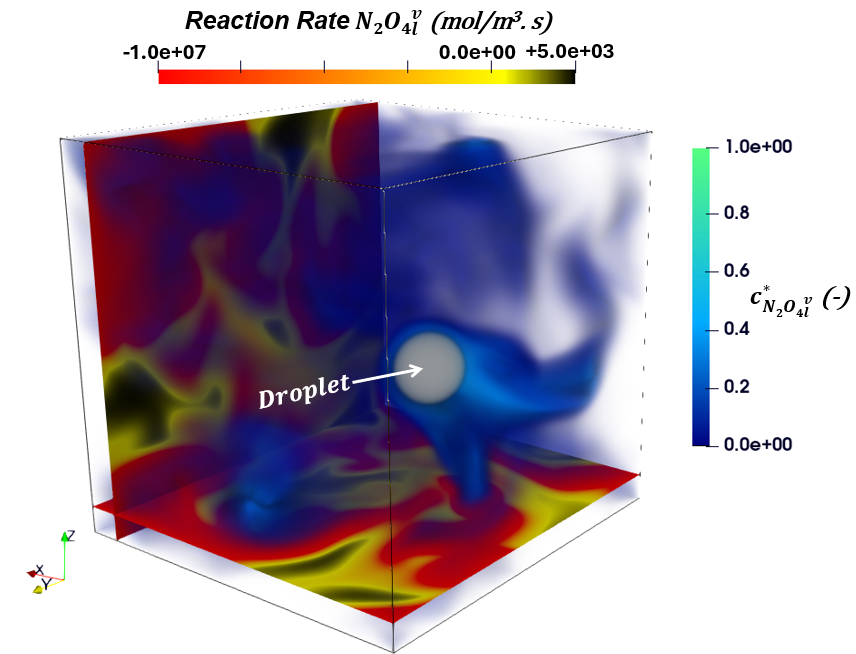}
        \caption{N\(_2\)O\(_4\)(g) normalized concentration}
        \label{fig:DNS_dns}
    \end{subfigure}
\caption{
(a) Instantaneous 3D contour plot depicting the spatial distribution of normalized evaporated \( \text{N}_2\text{O}_4 \) concentration at \( t = 200 \) ms, alongside the kinetics of the corresponding reaction rate.
\label{fig:DNS_turb}
}
\end{figure}

Finally, to examine the influence of turbulence on interfacial reactive processes by two scalar fields, we present a three-dimensional instantaneous contour plot in Figure~\ref{fig:DNS_dns}, obtained from direct numerical simulation using the ABC method. The simulation parameters include turbulent wave number (\( k_0 =2 \)) and turbulence flow amplitude (\( u_0=10 \)), facilitating interaction between phase transfer and reaction processes. Figure~\ref{fig:DNS_dns} illustrates the spatial distribution of normalized \(\text{N}_2\text{O}_4\) (g) concentration and the corresponding reaction rate at \( t = 200 \) ms. The color scales represent key physical quantities: the top scale indicates the reaction rate of N\(_2\)O\(_4\) (g) in mol/m\(^3\cdot\)s, where black and red regions denote exothermic and endothermic reactions. The right scale represents the normalized concentration of N\(_2\)O\(_4\) vapor, with blue indicating low concentration and green to cyan representing higher values. The gray spherical region marks the liquid-phase N\(_2\)O\(_4\) droplet, which serves as the source for phase transfer and reaction. This visualization highlights the complex coupling between turbulence, mass transfer, and chemical reactions, confirming the presence of exothermic and endothermic processes within the system.
\section{Conclusion}
This study presents a newly developed modeling tool to investigate the turbulent reactive phase change dynamics of N$_2$O$_4$(v,l) $\rightleftharpoons$ 2NO$_2$(v,l), utilizing Direct Numerical Simulation to capture the behavior of a dinitrogen tetroxide (N$_2$O$_4$) droplet undergoing reactive evaporation within a gaseous N$_2$O$_4$-NO$_2$ mixture. By employing Reactive Vapor-Liquid Equilibrium (RVLE) thermodynamics, we analyze the interplay between chemical reactions, phase transitions, and turbulence in heat and mass transfer. The results demonstrate that chemical reactivity significantly influences phase transition dynamics, altering critical points and thermodynamic saturation lines. Specifically, the forward and backward reactions play a crucial role in shaping the interfacial concentration gradients and phase boundary behavior in the reversible mechanism. While non-reactive phase transitions generally exhibit endothermic behavior, the reactive phase-change system reveals a dual thermal behavior, driven by the competing nature of the reactions. The dissociation of N\(_2\)O\(_4\) into NO\(_2\) is endothermic, absorbing heat and leading to a cooling effect at the liquid-vapor interface, which reduces local temperature and suppresses further evaporation. On the other hand, the recombination of NO\(_2\) into N\(_2\)O\(_4\) is exothermic, releasing heat and raising the temperature at the interface. This localized heating accelerates the evaporation rate, enhancing mass transfer. The dynamic interaction between these endothermic and exothermic effects results in a thermally dynamic interface, where the balance between heating and cooling shifts based on the relative rates of the forward and backward reactions, as well as the phase transfer. Finally, turbulence enhances mixing efficiency and accelerates the gas-liquid separation process, facilitating more efficient reactive mass transfer. These factors collectively influence system performance, providing insights into the role of reactive phase-change fluids in optimizing heat transfer processes. The findings highlight the importance of accounting for phase transition, chemical reactivity, and turbulence in the design of thermal systems with enhanced energy efficiency.

\section*{Acknowledgments}

The simulations and data management were carried out using resources provided by the National Academic Infrastructure for Supercomputing in Sweden (NAISS) at the PDC Center for High Performance Computing, with partial funding from the Swedish Research Council under grant agreement no. 2022-06725. Additional support was provided by the Swedish Research Council under grant agreement no. 2019-04086.
\bibliographystyle{refsTemplate}
\bibliography{refs}

\end{document}